\documentclass{aa}
\usepackage{graphicx}
\usepackage{txfonts}
\usepackage{subcaption}
 \usepackage{enumitem} 
 \usepackage[normalem]{ulem}
 \usepackage{orcidlink}
\usepackage{hyperref}
\hypersetup{colorlinks=true, citecolor=blue, urlcolor=teal}
\usepackage{orcidlink}

\usepackage{tikz}

\newcommand{\uca}{Universit\'e Clermont Auvergne, CNRS/IN2P3, LPCA, F-63000 Clermont-Ferrand, France}
\newcommand{\lyon}{Univ Lyon, Univ Claude Bernard Lyon 1, CNRS, IP2I Lyon/IN2P3, UMR 5822, F-69622, Villeurbanne, France}

\newcommand{\hmpc}{$\mathrm{Mpc.h^{-1}}$}
\newcommand{\hmpcvol}{$\mathrm{Mpc^{3}.h^{-3}}$}

\begin{document}

   \title{ZTF SN Ia DR2: Exploring SN Ia properties in the vicinity of under-dense environments}

   \author{M. Aubert
   \inst{\ref{uca}}\fnmsep\thanks{Corresponding author : marie.aubert@clermont.in2p3.fr}     
          \and P. Rosnet \inst{\ref{uca}}
          \and B. Popovic \inst{\ref{lyon}} \orcidlink{0000-0002-8012-6978} 
          \and F. Ruppin \inst{\ref{lyon}} \orcidlink{0000-0002-0955-8954} 
          \and M. Smith \inst{\ref{lancaster}} \orcidlink{0000-0002-3321-1432} 
          \and M. Rigault \inst{\ref{lyon}} \orcidlink{0000-0002-8121-2560}
          \and G. Dimitriadis \inst{\ref{dublin}} \orcidlink{0000-0001-9494-179X}
          \and A. Goobar \inst{\ref{physics_klein}} \orcidlink{0000-0002-4163-4996}
          \and J. Johansson \inst{\ref{physics_klein}}
          \orcidlink{0000-0001-5975-290X}
          \and C. Barjou-Delayre \inst{\ref{uca}}
          \and U. Burgaz \inst{\ref{dublin}} \orcidlink{0000-0003-0126-3999} 
          \and B. Carreres \inst{\ref{cppm},\ref{duke}} \orcidlink{0000-0002-7234-844X}
          \and F. Feinstein \inst{\ref{cppm}} \orcidlink{0000-0001-5548-3466}
          \and D. Fouchez \inst{\ref{cppm}} \orcidlink{0000-0002-7496-3796}
          \and L. Galbany \inst{\ref{barcelona1}, \ref{barcelona2}}\orcidlink{0000-0002-1296-6887}
          \and M. Ginolin \inst{\ref{lyon}} \orcidlink{0009-0004-5311-9301} 
          \and T. de Jaeger \inst{\ref{lpnhe}} \orcidlink{0000-0001-6069-1139} 
          \and M. M. Kasliwal \inst{\ref{caltech_dep}} \orcidlink{0000-0002-5619-4938}
          \and Y.-L. Kim \inst{\ref{lancaster}} \orcidlink{0000-0002-1031-0796}
          \and L. Lacroix \inst{\ref{physics_klein}, \ref{lpnhe}}
          \orcidlink{0000-0003-0629-5746}
          \and F. J. Masci \inst{\ref{ipac}} \orcidlink{0000-0002-8532-9395} 
          \and T. E. M\"uller-Bravo \inst{\ref{barcelona1},\ref{barcelona2}} \orcidlink{0000-0003-3939-7167} 
          \and B. Racine \inst{\ref{cppm}} \orcidlink{0000-0001-8861-3052}
          \and C. Ravoux \inst{\ref{uca}}
          \orcidlink{0000-0002-3500-6635}
          \and N. Regnault \inst{\ref{lpnhe}} \orcidlink{0000-0001-7029-7901} 
          \and R. L. Riddle \inst{\ref{caltech_op}}
          \and D. Rosselli \inst{\ref{cppm}} \orcidlink{0000-0001-6839-1421} 
          \and B. Rusholme \inst{\ref{ipac}} \orcidlink{0000-0001-7648-4142} 
          \and R. Smith \inst{\ref{caltech_op}} 
          \orcidlink{0000-0001-7062-9726}
          \and J. Sollerman \inst{\ref{astro_klein}} \orcidlink{0000-0003-1546-6615}
          \and J. H. Terwel \inst{\ref{dublin},\ref{nordicobs}} \orcidlink{0000-0001-9834-3439}
          \and A. Townsend \inst{\ref{berlin}}
          \orcidlink{0000-0001-6343-3362}.
          }

   \institute{\uca \label{uca},
         \and
            \lyon \label{lyon}, 
         \and 
         Department of Physics, Lancaster University, Lancs LA1 4YB, UK \label{lancaster} 
         \and
        School of Physics, Trinity College Dublin, College Green, Dublin 2, Ireland \label{dublin} 
        \and
        The Oskar Klein Centre, Department of Physics, Stockholm University, SE-10691 Stockholm, Sweden  \label{physics_klein}
         \and
         Aix Marseille Université, CNRS/IN2P3, CPPM, Marseille, France \label{cppm} 
         \and
         Department of Physics, Duke University, Durham, NC 27708, USA \label{duke} 
         \and        
        Institute of Space Sciences (ICE-CSIC), Campus UAB, Carrer de Can Magrans, s/n, E-08193 Barcelona, Spain \label{barcelona1} 
        \and
        Institut d'Estudis Espacials de Catalunya (IEEC), 08860 Castelldefels (Barcelona), Spain \label{barcelona2} 
        \and
        LPNHE, CNRS/IN2P3, Sorbonne Université, Université Paris-Cité, Laboratoire de Physique Nucléaire et de Hautes Énergies, 75005 Paris, France \label{lpnhe} 
        \and Division of Physics, Mathematics, and Astronomy, California Institute of Technology, Pasadena, CA 91125, USA \label{caltech_dep}
        \and
        IPAC, California Institute of Technology, 1200 E. California Blvd, Pasadena, CA 91125, USA \label{ipac}
        \and
        Caltech Optical Observatories, California Institute of Technology, Pasadena, CA  91125 \label{caltech_op}
        \and
        Department of Astronomy, The Oskar Klein Center, Stockholm University, AlbaNova, 10691 Stockholm, Sweden \label{astro_klein}
        \and
        Nordic Optical Telescope, Rambla José Ana Fernández Pérez 7, ES-38711 Breña Baja, Spain \label{nordicobs} 
        \and 
        Institut für Physik, Humboldt-Universität zu Berlin, Newtonstr. 15, 12489 Berlin, Germany \label{berlin}
        }

   \date{Received ---; accepted ---}

 
  \abstract
  {The unprecedented statistics of detected Type Ia supernovae (SNe Ia) brought by the Zwicky Transient Facility enables us to probe the impact of the Large-Scale Structure on the properties of these objects.}
  {The goal of this paper is to explore the possible impact of the under-dense part of the large-scale structure on the intrinsic SALT2 light curve properties of SNe Ia and uncover possible biases in SN Ia analyses.}
   {With a volume-limited selection of ZTF-Cosmo-DR2 Type Ia supernovae overlapping with the SDSS-DR7 survey footprint, we investigate the distribution of their properties with regard to voids detected in the SDSS-DR7 galaxy sample. We further use Voronoi volumes as proxy for local density environments within the large-scale structure.}
   {We find a moderate dependency of the stretch toward the localisation around the void centre and none when considering colour. The local Voronoi volumes mostly affect the fraction of low/high stretch supernovae.}
   {With the current statistics available, we consider that the impact of high or low local density environment can be considered as a proxy for the colour of the host galaxy. Under-dense environments should not cause any biases in supernova analyses.}

   \keywords{SN Ia -- large-scale structure --  cosmic voids.}

\authorrunning{M. Aubert}
   \maketitle
%
\section{Introduction}

Type Ia supernovae (SNe Ia) are commonly used as distance indicators in astronomy to map the Universe at cosmological scales. Due to the underlying physics phenomenon triggering such transient events -- white-dwarf thermonuclear explosions~\citep{Hoyle:1960zz} -- SNe Ia constitute standardizable candles. As such, they enabled the discovery of the accelerated expansion of the Universe in the late 90s \citep{1998AJ....116.1009R,1999ApJ...517..565P} and remain as of today a key probe to constrain deviations from the standard model of cosmology through the estimation of the Hubble constant \citep{Freedman_2021, Riess_2022}, or the Dark Energy equation of state \citep {Brout:2022vxf, DES:2024tys}.

The luminosity distance of SN Ia is estimated using the empirical Tripp relation~\citep{Tripp:1997wt} based on their light curve properties: (i) the broader the light curve shape the brighter is the SN Ia~\citep{Phillips:1993ng} and (ii) the redder the light curve the fainter is the SN Ia~\citep{1996ApJ...473...88R}. The basic standardization procedure corrects linearly those two effects to decrease the intrinsic dispersions of SN Ia absolute magnitudes at maximum luminosity. 

The current state-of-the-art of SN Ia standardization uses the SALT2 model \citep{Guy07, 2010A&A...523A...7G}, or its improved version SALT3 \citep{Kenworthy:2021azy}, to determine the two parameters entering the Tripp relation: the stretch (i) characterizing the broadening of the light curve and the colour (ii) to account for its redness. 
Nevertheless, beyond the two-term correction, an environmental dependency was observed, first with the mass of the host galaxy; the two-term corrected luminosity is brighter for SNe Ia in massive host galaxies \citep{Kelly:2009iy, SNLS:2010kps, 2010MNRAS.401.2331L}. By adding the mass-step term as an environmental dependency in the standardization procedure, the residual dispersion in the Hubble-Lema\^{\i}tre diagram is typically 0.12~mag \citep{Brout:2022vxf, Popovic:2023ltp, Rubin:2023ovl, DES:2024tys}.

Since the first evidence of SNe Ia brightness correlation with the stretch parameter, it has also been identified that the stretch parameter is connected to the host properties, such as the galaxy morphology~(\citet{1989PASP..101..588F, Hamuy:1996sq} and~\citet{Pruzhinskaya:2020rxg} for a more recent study) or other environmental tracers~\citep{2010MNRAS.406..782S, NearbySupernovaFactory:2018qkd, 2018ApJ...854...24K, 2019JKAS...52..181K}. Since galaxy properties evolve in time, \citet{2021A&A...649A..74N} also evidenced a redshift drift of the SNe~Ia stretch distribution. Besides the correlation between the stretch and environment, it also appears that the linear stretch-magnitude is broken \citep{DR2_stretch}.

As for the colour of SNe Ia, this property has also shown some weak correlation to its galaxy host environment \citep{2013ApJ...770..108C, 2021ApJ...913...49P, 2023MNRAS.519.3046K}. In addition, evidence of this SN~Ia parameter dependency w.r.t. redshift has emerged recently, as summarized in the  ZTF-DR2-Cosmo companion paper~\citep{DR2_Brodie}.

Beyond the host galaxy properties, a new question arises: is there a correlation between SN Ia properties and the large-scale structure of the Universe (LSS)?
Indeed, the LSS traced by matter shows a complex web of structures delineating filaments, walls, clusters and cosmic voids, all with very specific properties. 
Supernovae in general have been confirmed to reproduce the clustering of matter in the universe \citep{Tsaprazi2022}, however, the dependency of specific supernova properties toward these environments has not been extensively explored.
While a first study hinted toward such a correlation between the stretch of SNe Ia and their distance to galaxy clusters, the densest part of the large-scale structure~\citep{2024ApJ...961..185L,DR2_CL}, other extreme density environments are found within the cosmic web: cosmic voids.

Cosmic voids, in particular, are under-dense extended structures of the LSS within which, less matter and galaxies are observed on average. Since their discovery in the large-scale structure~\citep{1978ApJ...222..784G, 10.1093/mnras/185.2.357} and thanks to the increased statistics provided by large redshift surveys, voids have been detected \citep{Hoyle:1960zz, Hoyle2005, Pan2012, 2017ApJ...835..161M} and extensively studied in order to extract cosmological constraints using their size distribution \citep{Sheth2004, Pisani2015a, Contarini2019, 2022A&A...667A.162C}, cross-correlation function with galaxies \citep{2022MNRAS.516.4307W, 2022MNRAS.513..186A, 2022A&A...658A..20H, 2023A&A...677A..78R} and lensing signal \citep{Krause2013, SanchezC2017, 2023A&A...670A..47B}, see \cite{Pisani2019} and reference therein for a review. 

The analysis presented in this paper aims to study potential correlations between SN Ia light curve properties and the large-scale structure of the Universe, focusing on its under-dense parts. The analysis is based on the 2nd data release of SNe Ia from the Zwicky Transient Facility (ZTF), called the ZTF-Cosmo-DR2 sample \citep{DR2_DR} in the following. The paper structure goes as follows:  Section~\ref{sec-Data}  describes the SN Ia and cosmic voids sample selection and creation for our analysis. Section~\ref{sec-SNIa_within_voids} investigates SN Ia properties in the vicinity of voids. In Section~\ref{sec-SN Ia-Voronoi}, we present the repartition and properties of SNe Ia in regard to the local large-scale density provided by Voronoi volumes. Finally, we discuss the results in Section~\ref{sec-discussion} and conclude in Section~\ref{sec-conclusions}.
 
\section{Data samples}
\label{sec-Data}

We first describe the ZTF-Cosmo-DR2 sample, then we explain the creation process of the void catalogues at low redshift as well as the derived quantities used in this paper. Finally, we present the final samples that will be used in the paper. 

\subsection{The ZTF SNe Ia sample COSMO-DR2}
\label{subsec-DataSN}

The particularity of the ZTF survey \citep{Bellm_2019, Graham_2019} is to be able to scan its visible sky (the Northern sky above a declination of $-30 \deg$) more than once per night thanks to its large field-of-view camera of 47~deg$^2$ \citep{2020PASP..132c8001D}. 
The analysis of the observations \citep{2019PASP..131a8003M} ranging from 2018 to 2020, corresponding to the first stage of ZTF (ZTF-I), provides a unique data set -- the ZTF-Cosmo-DR2 -- of more than 3000 spectroscopically-classified SNe Ia in the nearby Universe, with cosmological redshift $z \leq 0.15$ \citep{DR2_DR}. The SN Ia classification was based on spectra acquisition from the SED-machine \citep[SEDm,][]{2018PASP..130c5003B,Rigault_2019, 2022PASP..134b4505K} for the most part with additional contributions discussed in \citet{DR2_Data}. SNe Ia were further classified into different subtypes e.g 'norm', 'pec-91bg', 'pec-91t', details concerning the classification of SNe Ia and further analysis are provided in \citet{DR2_Data, DR2_Subtype, DR2_HostSpec}. Out of this sample, we obtain a volume-limited sample of almost 1200 SNe Ia up to $z < 0.06$ \citep{DR2_simu}. The host galaxy redshift is available for 60\% of the volume-limited sample, the remainder of SN Ia redshifts being estimated from the SN Ia spectra, representing an unprecedented well-characterized sample of SNe Ia.

The light curves of this data set were fitted with the help of the SALT2.4 model \citep{Guy07,2010A&A...523A...7G, SDSS:2014iwm}, using the \citet{2023MNRAS.520.5209T} retraining, to extract their stretch ($x_1$) and colour ($c$) parameters used for their standardization \citep{DR2_Data}. This analysis is not aiming at computing corrected distance modulus as distance indicators for cosmological analysis, but trying to see if cross-correlations exist between SN Ia light curve properties ($x_1$ and $c$) and the large-scale structure of the Universe, focusing on cosmic void properties.

To this end, we will only consider the volume-limited SNe Ia within the redshift range $z = [0.02, 0.06[$ overlapping with the large-scale structure catalogue, galaxy and voids, defined below. This selection will be further detailed in Sect.~\ref{subsec-DataFinal}.

\subsection{SDSS DR7: void and galaxy samples}
\label{subsec-DataVoid}

In this subsection, we first present the source galaxy sample and the data selection used to identify cosmic voids. We then briefly describe the void identification process and the quality cuts applied to our catalogues. 

In order to define the large-scale environment, we make use of the DR7.2 NYU value-added galaxy catalogues (NYU-VAGC)\footnote{\url{http://sdss.physics.nyu.edu/vagc/lss.html}}\citep{2005AJ....129.2562B,2008ApJ...674.1217P} built from the seventh Data Release of the Sloan Digital Sky Survey collaboration (SDSS-DR7) \citep{2009ApJS..182..543A}. In particular, we make use of the \texttt{full0} galaxy catalogue, dedicated to large-scale structure studies. This catalogue includes all available objects in the spectroscopic footprint of the SDSS-DR7 survey.
From this sample, we first select the contiguous footprint area of the north galactic cap of $6960 \; \mathrm{deg^2}$ with an $80\%$ lower limit requirement for the completeness. The completeness indicates the number of galaxies whose redshift was measured out of the total number of sources observed in the photometric survey. 

We then select all the available galaxies in the redshift range $z = [0.018, 0.11]$ yielding 341433 objects. The selected redshift range largely encapsulates that of the first selection applied to the SNe Ia sample. The minimum redshift is defined to be wider than that of the SN Ia selection to mitigate possible boundary effects. Similarly, we deliberately set an arbitrarily high redshift as the upper boundary to remove any boundary effects in the vicinity of the most populated redshift range of the selected SN Ia sample.

To identify voids, we make use of the Revolver algorithm~\citep{Revolver2019}, similar to the VIDE algorithm~\citep{Sutter2014}, based on the underlying ZOBOV~\citep{Neyrinck2008} which applies a Voronoi tessellation scheme to the galaxy comoving coordinates. The Voronoi tessellation paves the space of a discrete distribution of points. Around each point (i.e. galaxy) is defined a Voronoi cell by tracing all the bisectors between the considered point and its surrounding neighbours. Each Voronoi cell therefore encloses all the nearest neighbouring space to a point. The resulting volume of the cell $V_c$, shaped as a polyhedron, provides an estimation of the local density $\rho_{loc} = 1/V_c$ in the vicinity of the point.

The void-finding process goes through three stages. 
First, the Revolver algorithm converts the positions to comoving coordinates, using a given Flat-$\mathrm{\Lambda CDM}$ fiducial cosmology. In this paper, the adopted fiducial cosmology is set to $\Omega_m = 0.31$ from here and on, and all distances will be quoted in \hmpc.

The algorithm adds fake galaxy positions within the holes of the survey footprint and at its angular and redshift edges to bind the volume to which the Voronoi tessellation is applied, acting as a buffer to prevent strong edge effects. The angular footprint is provided by a Healpix map. The latter is generated from the \texttt{mangle} polygons describing the footprint of the SDSS-DR7 survey \citep{2008MNRAS.387.1391S, 2004MNRAS.349..115H}.
The ZOBOV algorithm then implements the Voronoi tessellation, which is applied to all the tracers, including the generated buffer galaxies. 
In order to prevent the natural growth of the volumes with redshift and account for incompleteness in the survey, weight corrections are applied to the Voronoi volumes. The weights applied to these volumes depend on two quantities: the completeness, with which the volume will be decreased if the completeness is low, and the selection function which will correct for eventual variations of the number density of objects as a function of redshift.

From the resulting information brought by both Voronoi tessellation and member galaxy positions, one can then extract the basic properties of the voids: 

\begin{itemize}
    \item The Voronoi volume $V_c$ associated with each galaxy in the sample.
    \item Weighted Voronoi volume $V_{c,w}$ associated with each galaxy in the sample.
    \item Void radius: The void finder and subsequent void definition do not impart any shape to the detected voids. The void radius is therefore derived from the total volume of the $N_g$ Voronoi cells defining the void reduced to that of a spherical radius quantity~\eqref{eq:radius}: 
    \begin{equation}
    \centering
        r_v = \left( \frac{3}{4 \pi} \sum_i^{N_g} {V^i_c} \right)^{1/3}.
    \label{eq:radius}
    \end{equation}
     
    \item Void centre: The Voronoi volume-weighted barycentre~\eqref{eq:bary} of the member galaxies of the void:
    \begin{equation}
    \centering
        \mathbf{X_v} =  \frac{1}{\sum_i^{N_g} {V^i_c}} \sum_i^{N_g} V^i_c  \mathbf{X^i_g}.
    \label{eq:bary}
    \end{equation}
    
\end{itemize}

\begin{figure}
    \centering
    \includegraphics[scale=0.8]{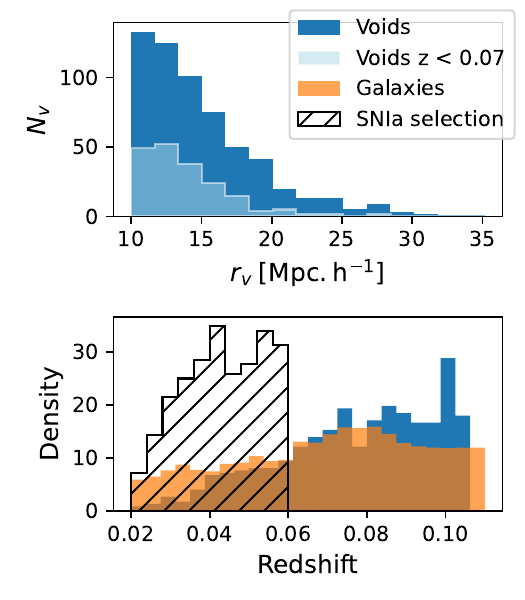}
    \caption{Summary statistics distribution. The top panel shows void radius distribution up to $ z < 0.11$ in blue and the void radius distribution with $z < 0.07$ in lighter blue. The bottom panel shows the void, galaxy and SN Ia redshift distributions.}
    \label{fig:gal-void-samp-stats}
\end{figure}

During the process of associating the Voronoi cells into voids -- see Fig. 1 of~\citet{Neyrinck2008} for a 2D illustration, galaxies adjacent to the buffer particles are flagged and taken into account in the construction of the void structures. Voids close to or associated with these flagged galaxies are considered edge voids and are discarded in our final selection to obtain a conservative void catalogue.
We apply a final conservative cut on the void radius: $r_v \geq 10$ \hmpc, corresponding to a value approaching $4 \times \mathrm{MPS}$. The MPS stands for mean particle separation, which is the average distance between galaxies in the sample and is about $2.5$\hmpc. This allows us to select voids that should be exempt from strong non-linear effects.
Figure~\ref{fig:gal-void-samp-stats} displays the void radius distribution, and the redshift distributions of both voids and galaxies. Our void sample contains in 592 voids up to a redshift of $z < 0.11$, which reduces to 193 voids with redshifts below $z < 0.07$.

\subsection{Matching SNe Ia and final data selection}
\label{subsec-DataFinal}

For the purpose of our study, we first select all the SNe Ia falling within the observed high completeness footprint of the SDSS DR7 NYU-VAGC. From this selection, we then consider only the volume-limited sample of SNe Ia by restricting our sample up to redshift $z < 0.06$.

All of these supernovae are then matched to their nearest neighbouring galaxy and voids through the use of a KD-Tree applied to their comoving coordinates, which provides the position of their nearest neighbouring object (void or galaxy) and their distance to that object. From this information, we discard all the SNe Ia in each sample whose nearest neighbouring galaxy is flagged as an edge galaxy. 

\begin{figure}
    \centering
    \includegraphics[width=\columnwidth, scale=0.8]{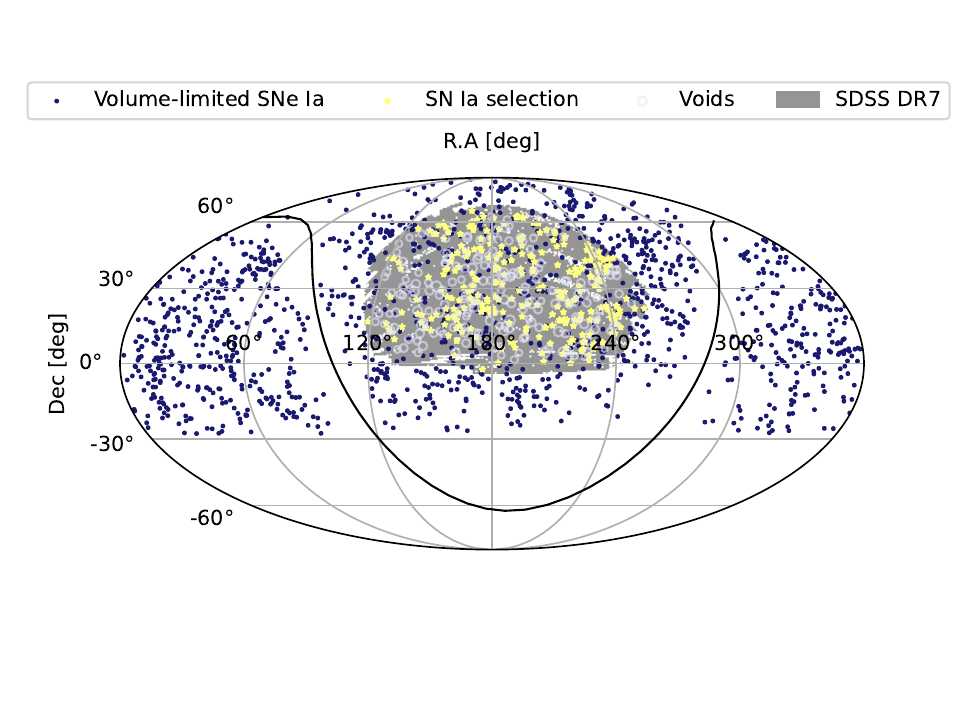}
    \caption{Angular distributions of the volume-limited ZTF-Cosmo-DR2 (dark blue dots), the reduced SN Ia sample after selection cuts (pale yellow stars) and voids (pale blue empty circles). For the latter, the marker sizes are scaled by the void radii. The black line corresponds to the galactic plane.}
    \label{fig:gal-void-samp-scatter}
\end{figure}

The total number of SNe Ia enclosed within the galaxy sample is therefore 448.
We further narrow down the SN Ia sample by applying similar quality cuts as detailed in \citet{DR2_stretch}. The considered cuts are the following: Good light curve sample with at least seven $5\sigma$ level detections within the $-10$ to $+40$ days phase range considered, among which $2$ pre- and post-max detections respectively and a detection in at least two of the ZTF bands (g, r, i), an absolute stretch parameter $|x_1| < 3$ with an error $\sigma_{x_1} < 1$, a colour $c \in [-0.2, 0.8]$ with an error less than $\sigma_c < 0.1$, an error on the estimated time of the peak luminosity below 1 day, $\sigma_{t_0} < 1$ day and finally, a SALT2 light curve fit probability greater than $10^{-7}$. 
Considering the quality cuts, the total number of SNe Ia considered in this paper, unless stated otherwise, is 280. Figure~\ref{fig:gal-void-samp-scatter} displays the SDSS-DR7 footprint, superimposed with angular positions of the original volume-limited SNe Ia sample, the reduced SNe Ia sample and the void sample. The considered counts for each tracer (SNe Ia, galaxies and voids) are quoted in Table~\ref{tab:recap_stat} and the resulting SN Ia redshifts distribution is shown in the lower panel of Fig.~\ref{fig:gal-void-samp-stats}. 

\begin{table}
    \centering
    \begin{tabular}{ccc}
        \textbf{Objects}  & \textbf{In SDSS} & \textbf{Quality cuts} \\ \hline
        SNe Ia & 448 & 280 \\
         \hline \hline
          & \textbf{Total} & \textbf{z < 0.07} \\ \hline
        Galaxies &  341433 &  157083 \\
        Voids & 592  & 193 \\
    \end{tabular}
    \caption{Summary number of each tracer. For SNe Ia, we quote the number within the SDSS sample, in the volume-limited prescription and after quality cuts. For galaxies, we quote the original selection and the $z < 0.07$ reduced count. For voids, we quote the final selection, including edge cuts and radius cuts, as 'Total' and the same sample with the added redshift cut $z < 0.07$.}
    \label{tab:recap_stat}
\end{table}

\section{SNe Ia within voids}
\label{sec-SNIa_within_voids}

In this section, we consider the repartition of the ZTF Type Ia supernovae within the SDSS footprint according to their distance from the centre of the nearest neighbouring voids. 

\begin{figure}[htb]
    \centering
    \includegraphics[scale=0.7]{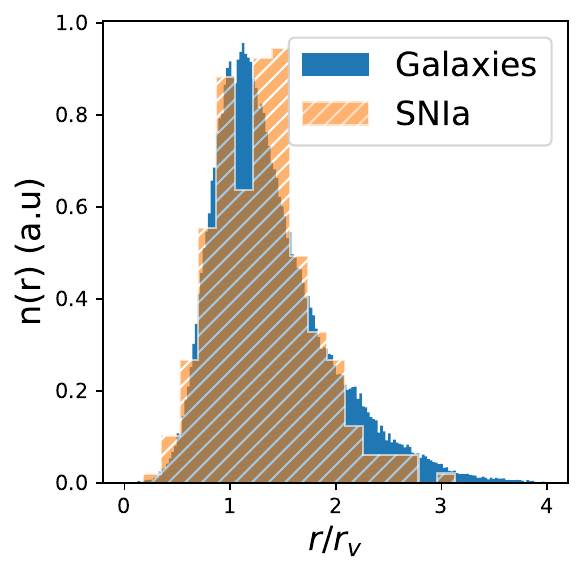}
    \caption{Void-centric distance of galaxies and SNe Ia distribution.}
    \label{fig:void-centric-dist}
\end{figure}

Using the matched void-SN pair mentioned above, we normalize the void-centric distance by the radius of the corresponding voids. The normalization allows us to be independent of the different scales of the void sample. Figure~\ref{fig:void-centric-dist} displays the corresponding distribution for the SNe Ia and galaxies matched with the same procedure. The SNe Ia distribution covers a similar distance range as that of galaxies and follows a similar behaviour, that is a low number of SNe Ia near the centre of the voids which then steadily increases and reaches a maximum value before lowering at higher separation $r/r_v > 2$. This suggests that SNe Ia also sample the large-scale structure, as galaxies do. We note however that the galaxy sample underlying the SNe Ia distribution might not be the same as the one probed by the SDSS sample.
Most SNe Ia in this sample are found within the range of 1 to 2 $R/R_v$ which corresponds to the characteristic compensating wall region surrounding voids and encapsulates the denser part of the large-scale structure such as clusters and filaments.

\begin{figure}
    \centering
    \includegraphics[width=\linewidth]{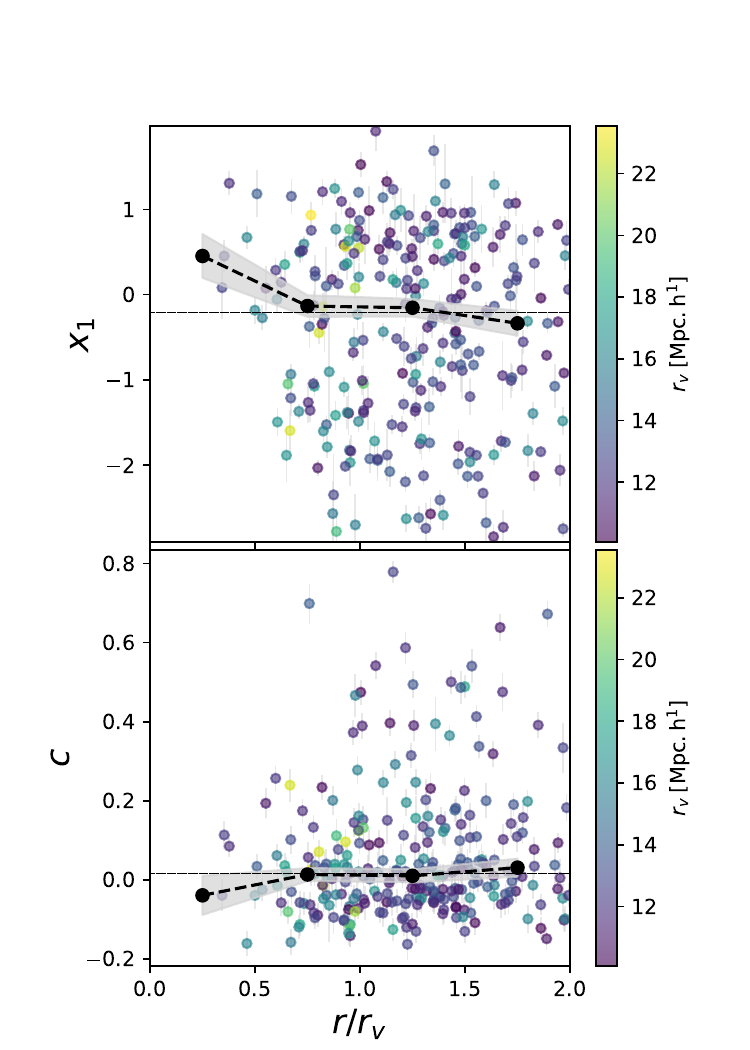}
    \caption{Scatter plots of the intrinsic properties of SN Ia as a function of their distance from the void centre, normalized by the void radius. The upper panel displays the stretch $x_1$ and the lower panel displays the colour $c$ of each SN Ia.
    The black dashed line with filled circle markers corresponds to the median value of the considered intrinsic property as a function of the midpoint of the $r/r_v$ bin. The shaded grey represents the standard deviation normalized by the count in the bin. The horizontal line displays the median value of the considered intrinsic parameter in this sample.
    The marker colour indicates the void radius $r_v$ in \hmpc of the nearest neighbouring void to the SN Ia.  The number of SNe Ia is the following: 5, 76, 116, 64, 19 in the five bins defined between in the range $r/r_v = [0,2]$.}
    \label{fig:prop_all}
\end{figure}

Figure~\ref{fig:prop_all} displays the scattered stretch (top main panel) and colour (bottom main panel) of each SN Ia according to its normalized distance from the centre of their nearest void. The dashed line shows the median in each bin to mitigate the effect of outlier values. 
Concerning the distribution of the stretch w.r.t. the distance to the void centre, the running median shows no significant trend for the stretch in the outer part of the void centre. The population of SNe Ia available below $0.8 r/r_v$ decreases significantly, and it seems that the number of low-stretch (negative) SNe Ia in this range decreases as well. The inner part of the void centre $< 0.5 r/r_v$ seems to favour positive stretch but it is to be noted that the number of SNe Ia in this region is $5$.
Regarding the colour, again, the running median shows no significant trend toward a favoured intrinsic colour regime within voids. It is  interesting however to note that the number of outlier values significantly decreases in the innermost part from the void centre.

Selecting a significant void sample with radii $\geq 10$ \hmpc, we found no significant behaviour of the stretch or colour as a function of the void-centric distance of the SN Ia. While this behaviour could be attributed to a lack of dependency of the SN Ia intrinsic properties w.r.t. the under-dense environment of the large-scale structure, some additional factors might come into play. The void definition used in this paper does not impose a shape (e.g., sphericity), as such, the void finding process might include higher-density regions on the surrounding of the voids in their definition. This probably results in a bias in the void centre definition and the radius. 

A second aspect to take into account is the lack of statistics at low $r/r_v$. Voids are, by definition, regions where the quantity of matter is lower than on average. Thus there are fewer galaxies inside voids and subsequently much less SNe Ia ~\citep{Tsaprazi2022}. One of the consequences is that the current SNe Ia sample might not allow us to sample the under-dense regions with high accuracy. 

\section{SNe Ia and Voronoi volume information}
\label{sec-SN Ia-Voronoi}

In this section, we consider the repartition of the ZTF type Ia supernova properties within the SDSS footprint according to the Voronoi volume of their nearest neighbouring galaxy.
As explained in Sect.~\ref{subsec-DataVoid}, the void finding process uses Voronoi tessellation to estimate the local density around each galaxy to proceed to the identification of low-density regions. The Voronoi volume is, by definition, indicative of the local density in the cosmic web: large (small) Voronoi volumes correspond to locally low (high) density regions of the large-scale structure.

This quantity has already been used as a marker of isolation of galaxies in simulations (e.g. in \citet{Habouzit2020}) to test their properties' dependency on the underlying density field. Additionally, following from the section above, the void-centric distribution of the SN Ia intrinsic properties might be biased because the defined voids are not spherical and might encapsulate further selection effects such as the inclusion of high-density galaxies in the ridge of the void~\citep{2024arXiv240320008Z}. 
The Voronoi volume therefore allows us to have an additional local density estimate within the LSS while avoiding potential systematic effects involved in the definition of voids.

\begin{figure}
    \centering
    \includegraphics[width=0.8\linewidth]{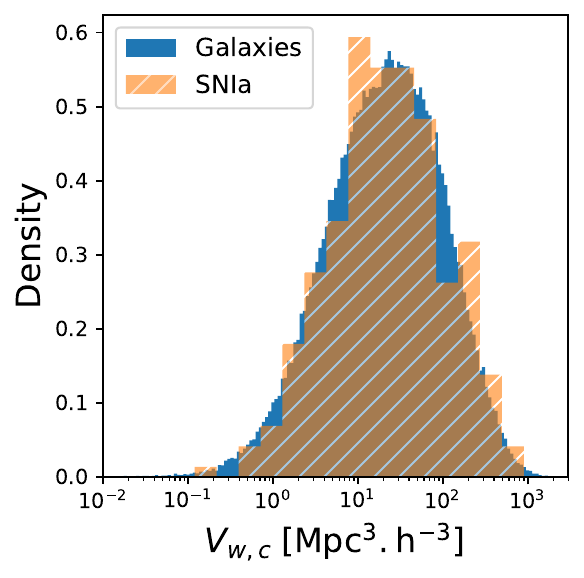}
    \caption{Voronoi volume distributions of the galaxy sample and matched SNe Ia.}
    \label{fig:Volume-dist}
\end{figure}

We make use of the matching of each supernova to its nearest neighbouring galaxy that was performed to discard SNe Ia too close to the edges of the survey. Then, each supernova is associated with the weighted Voronoi volume $V_{w,c}$ corresponding to its paired galaxy. We use the weighted volumes, thus corrected for completeness and the galaxy selection function. This analysis choice allows us to alleviate the correlation between the average Voronoi volume at a given redshift and the redshift of the matched galaxy. This also prevents us from probing the redshift drifts of SNe Ia \citep{2021A&A...649A..74N, DR2_Brodie} as we investigate the relation between Voronoi volumes and the matched supernova intrinsic properties.
Figure~\ref{fig:Volume-dist} shows the distributions of the Voronoi volume of the galaxy sample and the resulting distribution matched to the SNe Ia. Similar to the void-centric distance, SNe Ia do not seem to favour a volume range different from that of the galaxies.

\subsection{SN Ia properties according to Voronoi volumes}

As in the previous section, we investigate a possible dependency of the intrinsic properties, the stretch $x_1$ and the colour $c$, according to the weighted Voronoi volume of their matched galaxy.
We define six bins of Voronoi volumes in \hmpcvol from the minimum Voronoi volume matched to a SN Ia to its maximum to map the different local densities across the range of available volumes, where the last (biggest) volume bin is considered to probe the \textit{most} under-dense environment. %

\begin{figure}[htb]
    \includegraphics[width=\linewidth]{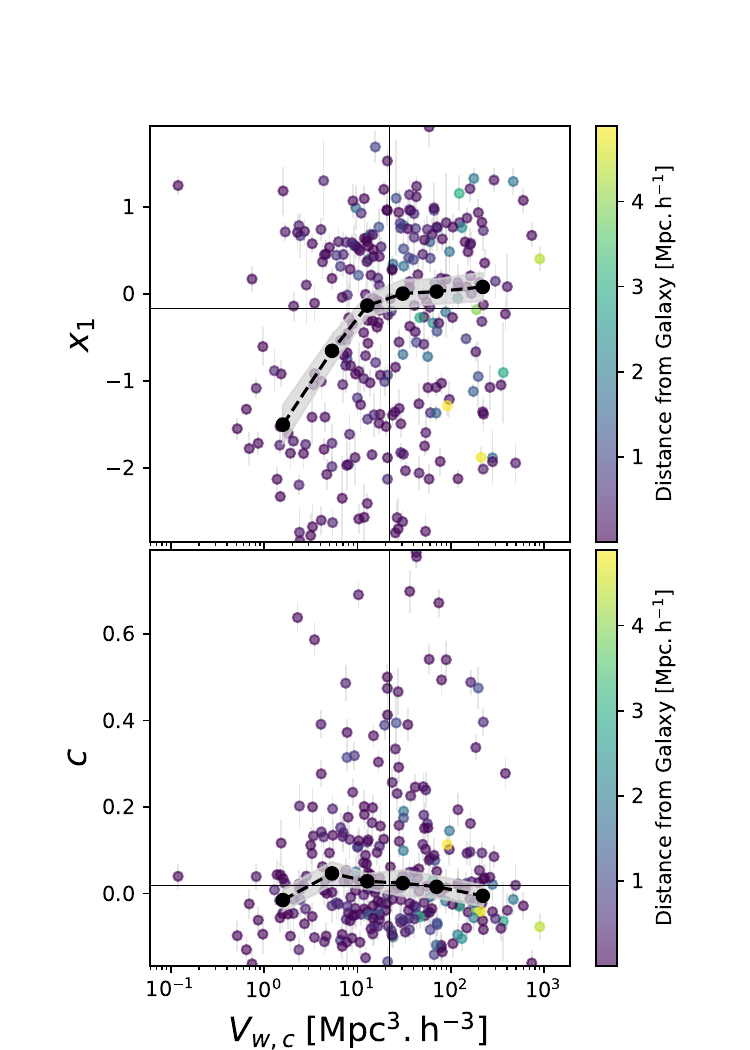}
    \caption{Scatter plots of the intrinsic properties of SN Ia as a function of the Voronoi volume in $\mathrm{Mpc^3.h^{-3}}$ of the nearest galaxy to the SN Ia. The upper panel display the stretch $x_1$ and the lower panel displays the colour $c$ of each individual matched SN Ia.
    The black dashed line with filled circled markers corresponds to the median value of the considered intrinsic property in each bin of volume, the volume values are taken to be the centroidal median values in each bin. The shaded grey represents the standard deviation normalized by the count in the bin. The number of SNe Ia in each bin, in ascending order of volumes, is the following: 30, 40, 65, 57, 49 and 39. 
    Vertical lines represent the median weighted Voronoi volume of the galaxies, $\bar{V}^{w}_c = 21.80$ \hmpcvol, and horizontal lines represent the median value of the considered intrinsic parameter in the sample.
    The marker colour indicates the distance in \hmpc to the nearest neighbouring galaxy to the SN Ia.}
    \label{fig:properties-voro-ag-redux}
\end{figure}

Figure~\ref{fig:properties-voro-ag-redux} displays the scattered stretch (top main panel) and colour (bottom main panel) of each SN Ia according to the Voronoi volume of their nearest galaxy. The dashed line shows the median stretch in each bin to mitigate the effect of outlier values. 
In the stretch distribution w.r.t. the matched Voronoi volumes, a strong increase of the stretch can be seen as the volume grows up until the median Voronoi volume of the galaxy sample, $\bar{V}^{w}_c = 21.80$ \hmpcvol, is reached (vertical dashed line). This trend is then suppressed after crossing into the sparser (less dense) region of Voronoi volume, which we defined to be $V^{w}_c > \bar{V}^{w}_c$. A relation can therefore be drawn between SNe Ia with negative stretch and high-density regions of the large-scale structure. 
In the case of the colour of the supernovae, no dependency can be seen.

\begin{figure}
    \centering
    \includegraphics[scale=0.7]{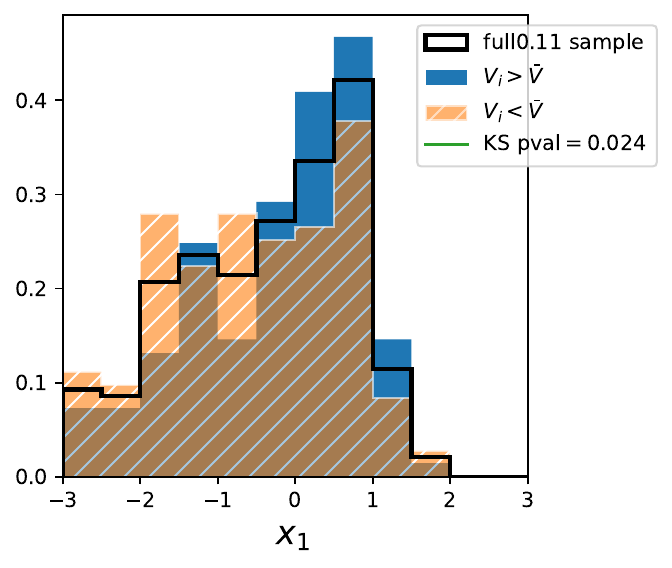}
    \caption{Histograms of the stretch in the full sample (black steps), low-density environment sample (blue histogram) and high-density environment sample (orange histogram). The p-value for the Kolmogorov-Smirnoff test applied to the low and high-density environment is provided in the legend and implies a significant difference.} 
    \label{fig:stretch-dist-volcut}
\end{figure}

Following the observed trend, we separate the SN Ia stretch sample into two distinct subsamples: the over-dense subsample is defined by SNe Ia with a matched volume below the median value of the galaxy sample and the under-dense subsample is defined by SNe Ia with a matched volume above the median value of the galaxy sample. 
Figure~\ref{fig:stretch-dist-volcut} shows the resulting stretch distributions of the two selected samples, as well as that of the full SN Ia sample used in this analysis. The Kolmogorov-Smirnoff test applied to the two subsamples returns a p-value of 0.024, meaning that there is only a 2.4$\%$ probability that the samples originate from the same distribution. 
We can see that the low stretch end is more populated in the over-dense bin than in the under-dense one compared to the complete sample distribution.

We can now attempt to understand exactly how the Voronoi volumes affect the stretch distribution. To do so, we suppose the probability distribution function (PDF) of the stretch distribution is bimodal \citep{2020A&A...644A.176R, 2021A&A...649A..74N} and can be described as a simple mixture of two Gaussian distributions:
\begin{equation}
    \mathcal{P}(x_1) = r \mathcal{N}(\mu^{high}_{x_1} , \Tilde{\sigma}^{high}_{x_1}) +  (1-r) \mathcal{N}(\mu^{low}_{x_1} , \Tilde{\sigma}^{low}_{x_1}),
\end{equation}
where $\mathcal{N}$ is a normal distribution, characterized by its mean stretch $\mu_{x_1}$ and standard deviation $\Tilde{\sigma}_{x_1}$. $r$ is the fraction of SNe Ia belonging to the high-stretch mode.
The $\Tilde{\sigma}_{x_1}$ is defined as follows:
\begin{equation}
\centering
{\Tilde{\sigma}^{high/low}_{x_1}\;}^2 = {\sigma^{high/low}_{x_1}\;}^2 + {\Delta_{x_1}\;}^2,
\end{equation}
where $\Delta_{x_1}$ is the error of the measured stretch.

\begin{table}
    \centering
    \begin{tabular}{cccc}
       Sample  & Full & over-dense & under-dense \\
       \hline \\
        $\mu_{x_1}^{low}$ & $-1.20 \pm 0.21$ &  $-1.39 \pm 0.17$ & $-1.52 \pm 0.22$  \\
         $\sigma_{x_1}^{low}$ & $0.74 \pm 0.13$ &  $0.83 \pm 0.14$  &  $0.62 \pm 0.15$ \\
        $\mu_{x_1}^{high}$ &  $0.45 \pm 0.07$  &  $0.59 \pm 0.11$  & $0.38 \pm 0.09$ \\
        $\sigma_{x_1}^{high}$ & $0.48 \pm 0.07$ & $0.36 \pm 0.10$ &  $0.53 \pm 0.07$\\
         $r$ & $0.54 \pm 0.08$   & $0.37 \pm 0.09$   &  $0.69 \pm 0.07$  \\
    \end{tabular}
    \caption{Best-fitting parameters to the two Gaussian mixtures PDF of the stretch distribution.}
    \label{tab:bimodal}
\end{table}

We fit this parametric model to the total sample and the subsamples, the resulting values are quoted in Table~\ref{tab:bimodal}. 
The best-fitted values to the full sample agree within $1\sigma$ to those fitted in~\citet{DR2_stretch}. When comparing the over-dense and under-dense cases quoted in the table, we can see that the parameter most affected by the split into two volume categories is the fraction of SNe Ia in the high stretch mode $r$. We reach a $2.76\sigma$ significance of the difference of the best-fitting fractions. The best-fitted mean stretch values in the low and high modes differ with no and moderate significance of $1.08\sigma$ and $1.48\sigma$ respectively.
The best-fitted standard deviation of the high mode also reaches a $1.46\sigma$ difference. This can be explained from Fig.~ \ref{fig:properties-voro-ag-redux} where we can see that the dense stretch distribution has a much more defined bimodal shape with a sharp high stretch mode. In the under-dense bin case, the transition between the two Gaussian mixtures seems smoother. 

\subsection{SN Ia repartition according to Voronoi volumes}

\begin{figure*}[htb]
    \centering
    \includegraphics[width=0.8\textwidth]{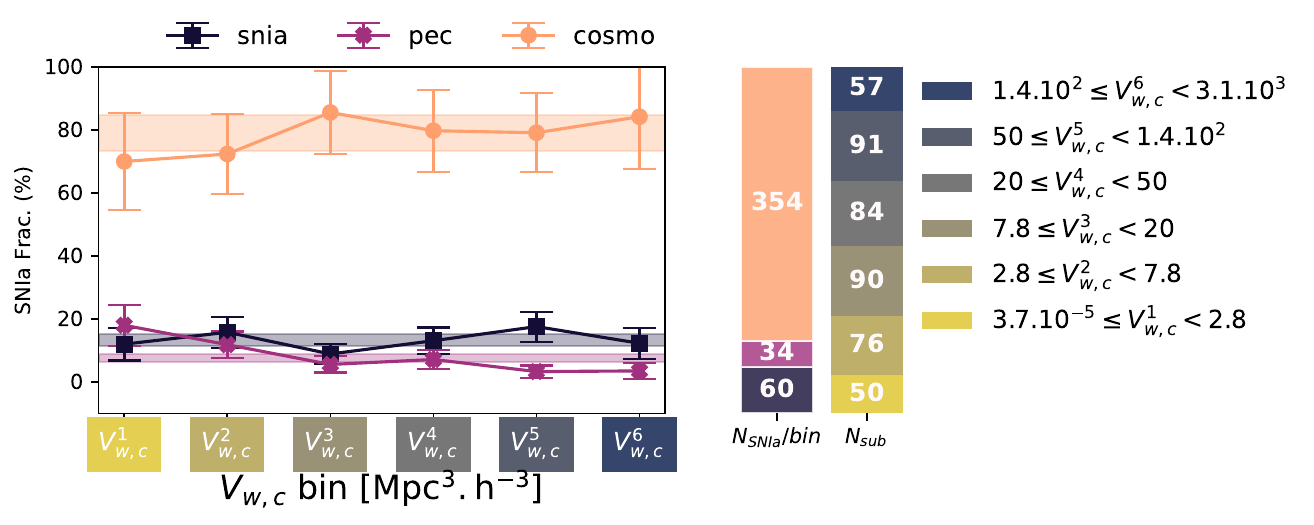}

    \caption{SN Ia subtype fractions as a function of the binned nearest galaxy volumes in $\mathrm{Mpc^3.h^{-3}}$. The shaded horizontal bands correspond to the fractional repartition of the SNe Ia subtypes in the sample, corresponding to the first stacked vertical histogram on the right. In the latter, we also superimpose the number of SNe Ia in each subtype category. The outer right vertical histogram shows the fractional number of SNe Ia in each volume bin, with the superimposed number of SN Ia considered in this bin.}
    \label{fig:SNIa-vol-rep}
\end{figure*}

Up until now, we have been considering all SNe Ia as being of the same type. We now consider a possible dependency of these subtypes according to their local density environment, as probed by the Voronoi volumes. The resulting counts of SNe Ia according to the volumes of their matched galaxies and the subtypes' repartition are represented in Fig.~\ref{fig:SNIa-vol-rep}.  The subtypes are distributed into three broad categories:\texttt{cosmo}, \texttt{snia} and \texttt{peculiar} (pec). The \texttt{cosmo} category encapsulates most norm or 91t SNe Ia, \texttt{snia} categorizes SNe Ia that cannot be classified as norm with certainty and \texttt{peculiar} account for non-norm or non-91t SNe Ia, further details on this classification process and analyses on SNe Ia subtypes can be found in the ZTF companion papers \citet{DR2_Subtype, DR2_HostSpec}. 

To perform this simple check, we open up the selection of our data to all 448 SNe Ia found within the SDSS survey, as presented in Sect.~\ref{subsec-DataFinal}. This is necessary as most of the SNe Ia subtyped as peculiar do not pass the quality cut applied to our analyses.

We show the fraction of SNe Ia according to their types in each of the volume bins, from the densest part of the sample to the sparsest in Fig.~\ref{fig:SNIa-vol-rep}, as well as the corresponding histogram counts of the SNe Ia according to their types (the first stacked histogram) and their matched volume (the second stacked histogram). The error bars are computed by considering the Poisson counts in each bin.
About $79\%$ of the SNe Ia are classified as \textit{cosmo}, $7.6\%$ as \textit{peculiar} and $13.4\%$ as \textit{SN Ia}. The repartition of those subtypes in the different local environments is in general accordance with that of the total sample, within $1\sigma$, except for the population of \textit{peculiar} SN Ia in the densest bin $V^1_{w,c}$, accounting for $17\%$ of the SNe Ia detected in this range, which represents a $1.7\sigma$ difference from the SN Ia fraction computed over the totality of our sample. 
These \textit{peculiar} SNe Ia in the densest bin are mostly \textit{91bg} types.
Considering these statistics, we can conclude that there is a very moderate dependency of the SN Ia subtypes on their local density environments. This moderate dependency is due to \textit{peculiar} SNe Ia that appear to be more prevalent in over-dense environments.
When considering the quality cuts applied to obtain a cosmological SNe Ia sample, this dependency disappears as most of the \textit{peculiar} do not pass these cuts.

\section{Discussion}
\label{sec-discussion}

Throughout this paper, we have investigated the relationship between the large-scale environment of SN Ia and their properties, with a particular focus on voids and overall under-dense environments. 
We first considered the environment traced by voids identified in the SDSS-DR7 galaxy sample and then extended the analysis toward using the Voronoi volumes of the galaxies, a by-product of the void finding process, as a tracer of local density in the large-scale structure.

Most SNe Ia are found at a distance of $r/r_v > 0.5$ which attests to their prevalence in medium to high-density environments such as filaments, walls and clusters. This observation also confirms that SNe Ia sample the galaxy density field, having a similar radial density, in accordance with \citet{Tsaprazi2022} clustering evidence. We observe $1\%$ of SNe Ia in the selected sample to be found in highly under-dense environment $r/r_v < 0.5$, while \citet{Tsaprazi2022}, with a sample of $498$ SNe Ia found about $4\%$ of those to be embedded in voids detected in density field reconstruction. We therefore detect less SNe Ia near voids than they do which can be attributed to the void definition procedure, ours being geometrical while theirs is dynamical. We still find consistent fractions between both galaxy  and SN Ia samples in the under-dense part of the voids. 
Beyond the similar clustering behaviour of SNe Ia around voids, we detect no significant behaviour of the SN Ia properties w.r.t. the normalized void-centric distance of the SNe Ia. However, this lack of evidence cannot be solely attributed to a lack of colour and especially stretch dependency in the void vicinity due to additional effects that might come into play. Voids are, by definition, lowly populated environments, so we might be limited by the SN Ia statistics especially in the range $[0 - 1] r/r_v$ of the void centre to detect decidedly a possible relation. A second statistical aspect might be the lack of void statistics, limited by the sky coverage of the source galaxy sample. 
Additionally, while some stringent cuts were applied to mitigate possible \textit{spurious} voids, other systematic effects might be coming into play in the void definition, such as off-centring or over-estimation of the radii. The definition from the Voronoi volume provides local density information, as such the defined voids might be part of an overall denser or more under-dense part of the universe. Some of these effects might be averaged out in the statistical two-point analyses but not with the methodology adopted in this paper.

When considering the local environment provided by the weighted Voronoi volumes, we find a strong dependency between the stretch and the aforementioned quantity: the denser the environment, the lower is the median stretch. This behaviour is in full agreement with the cluster-centric distribution of the stretch pointed out by \citet{2024ApJ...961..185L} and shown in the \citet{DR2_CL} ZTF-DR2 companion paper. 
Using the Voronoi volumes, we can trace a full evolution from the highest-density regions of the sample to the lowest. The volumes, in comparison to the cluster matching in the \citet{DR2_CL} analysis, enable us to categorize the SN Ia large-scale environment and access locally dense environments, such as dense galaxy groups or undetected clusters. Indeed, the \citet{DR2_CL} study points out that the cluster sample is incomplete for masses $M_{500} \lesssim 10^{13} M_{\odot}$ at $z \leq 0.1$. Voronoi volumes thus provide complementary information for the less dense environment than those galaxy clusters.
The influence of the large-scale structure seems stronger in high-density environments than in low-density environments when considering the scattered distribution. There can be several reasons for this behaviour. Firstly, high-density regions, therefore small volumes, are better sampled than large volumes (low-density regions) because they are inhabited by more luminous objects. Secondly, volumes in the middle to high volume range can, inversely, be less well sampled, especially at high redshifts where faint objects are not detected. 
By modelling our stretch distribution with a bimodal model, we found that the Voronoi volume affects the fraction of SNe Ia belonging to the high stretch mode. The decrease in the fraction of SNe Ia populating the high stretch mode in locally dense environments is in accordance with the model considered by \citet{DR2_CL} in the vicinity of clusters.

Although the stretch behaviour seems to depend on its local density environment, one can (legitimately) question whether this apparent dependency is actually originating from the large-scale structure not from host galaxy properties for which the Voronoi volume might be sensitive to.

Among the known environmental dependencies of SNe Ia, we refer to the local colour and global mass as used in \citet{DR2_stretch}, wherein they showed that the stretch distribution can be affected in two different ways, w.r.t. to the environmental tracer. The local colour affects mostly the fraction of SNe Ia in the high stretch mode, while the global mass shifts the average value of the stretch mode. 
In comparison, the apparent shift -- $1\sigma$ for the low stretch mode and above $1\sigma$ for the high stretch mode -- toward lower stretch values in the under-dense volume bin is in opposition to the behaviour highlighted in~\citet{DR2_stretch}. 
This observation pushes us toward discarding our density environment tracer as a direct proxy for the global mass of the host galaxies. However, we note that the behaviour of the best-fitting parameters might also be due to the smoother transition between the low and high stretch mode in the under-dense bin than in the over-dense bin which displays a much stronger bimodality.
 \citet{DR2_CL} discarded this hypothesis as well by considering a similar parametrization of the stretch mean values as~\citet{DR2_stretch}. 
However, in both our analysis and that of \citet{DR2_CL}, the density environment affects mostly the fraction of SNe Ia in the high stretch mode, thus having a similar effect as the local colour of the host galaxy. 
Works related to the study of galaxy properties w.r.t. their local environment evidenced a colour-density relation between the colour of galaxies and the local environment of those galaxies within the large-scale structure \citep{2009MNRAS.393.1324B,2004MNRAS.353..713K}, with redder galaxies being found in denser environments and bluer galaxies in less dense environments. 
This relation is further supported as galaxies within voids are less massive, bluer and favour younger stellar environments with high specific star formation rates \citep{Rojas2004, Ricciadelli2014, Ricciardelli2017, Florez2021, Habouzit2020}. The latter properties correspond to host galaxies of high stretch mode supernovae \citep{NearbySupernovaFactory:2018qkd, 2021A&A...649A..74N}. 

In light of these pieces of information, we can consider that the probable dependency on the local density environment is most likely linked to specific galaxy properties such as colour. Under-dense environments should favour high stretch SNe Ia which already consist in the dominant part of the signal, which could explain why no significant effect can be seen when considering Voronoi volumes.

\section{Conclusion}
\label{sec-conclusions}

Using a selection of 280 SNe Ia from the ZTF-COSMO DR2 sample overlapping with the Main Sample footprint of the SDSS galaxy survey, we investigated the relationship between SN Ia properties, stretch and colour, w.r.t. their distance to the centre of their nearest neighbouring voids. We identified voids within the low redshift $z < 0.1$ galaxy spectroscopic sample from the SDSS-DR7 data.
Beyond confirming that SNe Ia sample the density field in a similar manner as the SDSS galaxy distribution, we found no significant behaviour of the stretch or colour distribution w.r.t. the distance from the centre of the void, possibly due to lack of signal or additional effects such as low SN Ia sampling in the vicinity of the void or void selection. 

Using the Voronoi volumes of the nearest neighbouring galaxy to each supernova, we probed the dependency of stretch and colour with the local density of the SNe Ia. We found that small volumes, probing high-density regions, significantly favour low-stretch SNe Ia. 
Our results are consistent with the effect detected w.r.t. the cluster-centric stretch distribution, which favours low stretches in the vicinity of the cluster \citep{DR2_CL}. Volumes are also complementary as they allow us to be more sensitive to locally high-density regions which do not pass the detection threshold to be qualified as clusters, e.g. galaxy groups. 
This dependency affects the fraction of low-stretch / high-stretch supernovae in the sample and not their average distribution, similar to the host local colour, as shown by ~\citet{DR2_stretch}. Previous works also showed a relationship between the colour of galaxies and their local density in the large-scale environment. We thus consider that the dependency of SNe Ia to the large-scale structure might be mostly driven by the colour of their host galaxies. Both an increased coverage of the ZTF footprint by a low redshift galaxy catalogue and an increased SNe Ia sample might allow us to investigate this aspect further. 

The implication of such a relation should not incur strong biases in the analysis of the Hubble Diagram, provided that the treated SNe Ia sample is homogeneous and unbiased toward a specific large-scale environment. It might however be relevant in analyses driven toward density field and velocity field reconstruction, considering that SN Ia distance moduli might be loosely dependent on the spatial repartition of SNe Ia in the large-scale structure.

\begin{acknowledgements}
This work has been supported by the Agence Nationale de la Recherche of the French government through the program ANR-21-CE31-0016-03. This project has received funding from the European Research Council (ERC) under the European Union's Horizon 2020 research and innovation program (grant agreement n 759194 - USNAC). GD, UB and JHT are supported by the UB is supported by the H2020 European Research Council grant no. 758638. L.G. acknowledges financial support from AGAUR, CSIC, MCIN and AEI 10.13039/501100011033 under projects PID2020-115253GA-I00, PIE 20215AT016, CEX2020-001058-M, and 2021-SGR-01270. This work has been supported by the research project grant “Understanding the Dynamic Universe” funded by the Knut and Alice Wallenberg Foundation under Dnr KAW 2018.0067 and the  {\em Vetenskapsr\aa det}, the Swedish Research Council, project 2020-03444. Y.-L.K. has received funding from the Science and Technology Facilities Council [grant number ST/V000713/1]. T.E.M.B. acknowledges financial support from the Spanish Ministerio de Ciencia e Innovaci\'on (MCIN), the Agencia Estatal de Investigaci\'on (AEI) 10.13039/501100011033, and the European Union Next Generation EU/PRTR funds under the 2021 Juan de la Cierva program FJC2021-047124-I and the PID2020-115253GA-I00 HOSTFLOWS project, from Centro Superior de Investigaciones Cient\'ificas (CSIC) under the PIE project 20215AT016, and the program Unidad de Excelencia Mar\'ia de Maeztu CEX2020-001058-M.
Based on observations obtained with the Samuel Oschin Telescope 48-inch and the 60-inch Telescope at the Palomar Observatory as part of the Zwicky Transient Facility project. ZTF is supported by the National Science Foundation under Grant No. AST-2034437 and a collaboration including Caltech, IPAC, the Weizmann Institute of Science, the Oskar Klein Center at Stockholm University, the University of Maryland, Deutsches Elektronen-Synchrotron and Humboldt University, the TANGO Consortium of Taiwan, the University of Wisconsin at Milwaukee, Trinity College Dublin, Lawrence Livermore National Laboratories, and IN2P3, France. Operations are conducted by COO, IPAC, and UW. SED Machine is based upon work supported by the National Science Foundation under Grant No. 1106171. The ZTF forced-photometry service was funded under the Heising-Simons Foundation grant \#12540303 (PI: Graham). The Gordon and Betty Moore Foundation, through both the Data-Driven Investigator Program and a dedicated grant, provided critical funding for \href{https://www.oir.caltech.edu/twiki_ptf/bin/edit/ZTF/SkyPortal?topicparent=ZTF.SubtoBoard;nowysiwyg=0}{SkyPortal}. Funding for the Sloan Digital Sky Survey (SDSS) has been provided by the Alfred P. Sloan Foundation, the Participating Institutions, the National Aeronautics and Space Administration, the National Science Foundation, the U.S. Department of Energy, the Japanese Monbukagakusho, and the Max Planck Society. The SDSS Website is http://www.sdss.org/.
The SDSS is managed by the Astrophysical Research Consortium (ARC) for the Participating Institutions. The Participating Institutions are The University of Chicago, Fermilab, the Institute for Advanced Study, the Japan Participation Group, The Johns Hopkins University, Los Alamos National Laboratory, the Max-Planck-Institute for Astronomy (MPIA), the Max-Planck-Institute for Astrophysics (MPA), New Mexico State University, University of Pittsburgh, Princeton University, the United States Naval Observatory, and the University of Washington. 
\end{acknowledgements}

\bibliographystyle{aa} 
\bibliography{biblio} 

\end{document}